# Synthesis of superfine high-entropy metal diboride powders

Da Liu[1,a], Tongqi Wen[2,a], Beilin Ye[1], Yanhui Chu[1]*

[1]School of Materials Science and Engineering, South China University of Technology, Guangzhou 510641, China

[2]MOE Key Laboratory of Materials Physics and Chemistry under Extraordinary Conditions, School of Natural and Applied Sciences, Northwestern Polytechnical University, Xi'an 710072, China

**Abstract**

High-purity and superfine high-entropy metal diboride powders, namely $(Hf_{0.2}Zr_{0.2}Ta_{0.2}Nb_{0.2}Ti_{0.2})B_2$, were successfully synthesized via a facile borothermal reduction method at 1973 K for the first time. The as-synthesized powders with an average particle size of ~ 310 nm had a single-crystalline hexagonal structure of metal diborides and simultaneously possessed high compositional uniformity from nanoscale to microscale. In addition, their formation mechanisms were well interpreted by analyzing the thermodynamics of the possible chemical reactions based on the first principles calculations. This work will open up a new research field on the synthesis of high-entropy metal diboride powders.

---

[a]D. Liu and T.-Q. Wen contributed equally to this work.
* Corresponding author. Tel.:+86-20-82283990; fax:+86-20-82283990.
  E-mail address: chuyh@scut.edu.cn (Y.-H. Chu)

**Graphical Abstract**

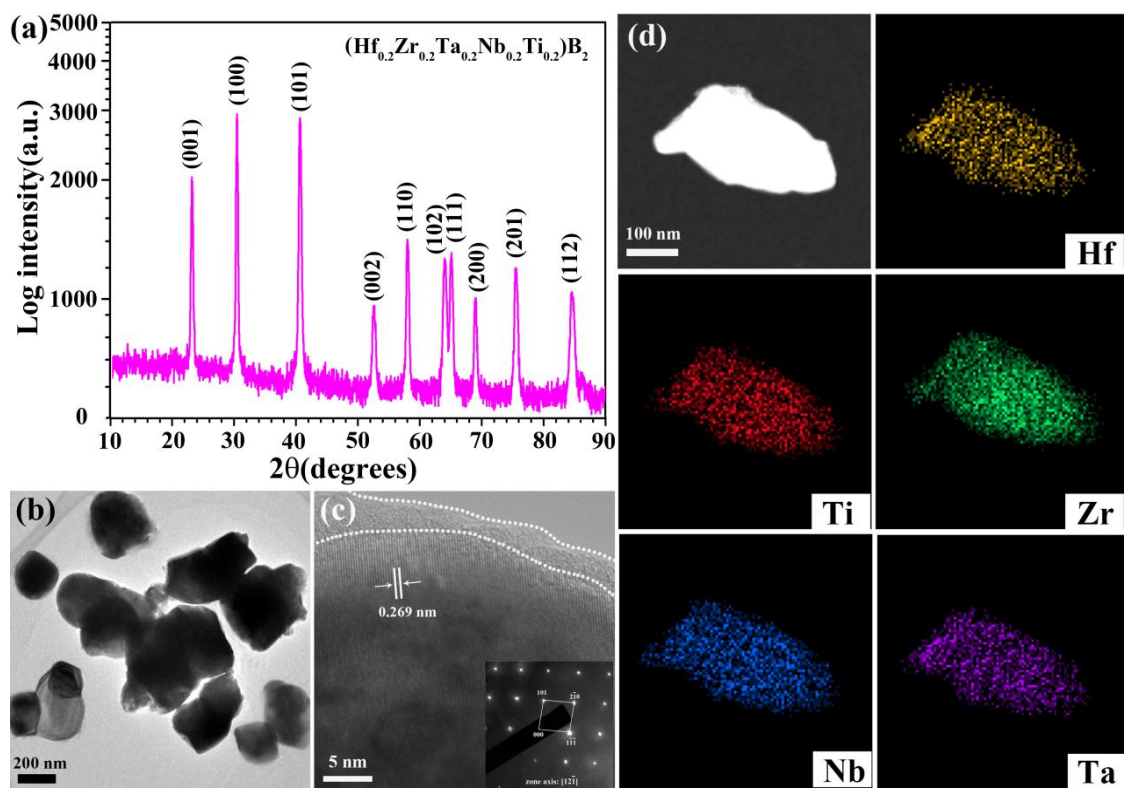

High-purity and superfine high-entropy metal diboride powders with a single-crystalline hexagonal structure and high compositional uniformity, namely $(Hf_{0.2}Zr_{0.2}Ta_{0.2}Nb_{0.2}Ti_{0.2})B_2$, were successfully synthesized via a facile borothermal reduction method at 1973 K for the first time.

Crystalline high-entropy ceramics, a new class of solid solutions that contain four or more principal metallic components in near-equiatomic ratios, are attracting increasing interest for their unique structure and properties and potential applications [1,2]. In recent years, a variety of crystalline high-entropy ceramics has been fabricated and studied, such as metal carbides [3-8], oxides [9-11], diborides [12,13] and nitrides [14]. Among these crystalline high-entropy ceramics, the high-entropy metal diborides have recently attracted significant attentions for potential applications in various fields, such as aerospace, solar energy, nuclear reactors, metallurgy, cutting tools, microelectronics, etc., in that their metal diboride components possess unusual combination of attractive physico-chemical properties, such as melting temperature exceeding 3000 K, high hardness, chemical inertness, good electrical and thermal conductivity, intrinsic solar selectivity, low neutron absorption, etc. [12,13,15]. For example, Gild *et al.* [12] fabricated six types of high-entropy metal diboride monolithic ceramics with five components by sparkle plasma sintering (SPS) technique. Tallarita *et al.* [13] prepared $(Hf_{0.2}Mo_{0.2}Ta_{0.2}Nb_{0.2}Ti_{0.2})B_2$ monolithic ceramic by a two-step technique consisting of self-propagating high-temperature synthesis and SPS. However, the presence of high porosity, large grain size, and low compositional uniformity severely weakens the performances of the high-entropy metal diboride monolithic ceramics. The synthesis of the high-entropy metal diboride powders is a key technique for addressing these issues. Nevertheless, to our best knowledge, the synthesis of the high-entropy metal diboride powders has never been reported until now.

In this work, the high-purity and superfine high-entropy metal diboride powders, namely $(Hf_{0.2}Zr_{0.2}Ta_{0.2}Nb_{0.2}Ti_{0.2})B_2$ (HMD-1), with high compositional uniformity were successfully synthesized by a simple borothermal reduction method at 1973 K for the first time. The effect of synthesis temperatures on the phase compositions of the as-synthesized powders was first studied. Subsequently, the morphology, crystal structure and compositional uniformity of the as-synthesized powders were investigated in detail, as well as the formation mechanism.

The commercially available $HfO_2$, $Nb_2O_5$ and $Ta_2O_5$ powders (purity: 99.9%, particle size: 1 ~ 3 μm, Shanghai ChaoWei Nanotechnology Co. Ltd., Shanghai, China), $ZrO_2$ and $TiO_2$ powders (purity: 99.9%, particle size: 100 ~ 300 nm, Shanghai ChaoWei Nanotechnology Co. Ltd., Shanghai, China), and amorphous B powders (purity: 99.9%, particle size: 1 ~ 3 μm, Shanghai ChaoWei Nanotechnology Co. Ltd., Shanghai, China) were used as the starting materials. Details of the synthesis of HMD-1 powders using borothermal reduction method were described as follows: The starting materials were first milled by ball milling with $ZrO_2$ balls for 6 h using the following molar ratios: 2:2:2:1:1 for $ZrO_2/HfO_2/TiO_2/Ta_2O_5/Nb_2O_5$ and 2:11 for $(HfO_2, ZrO_2, Ta_2O_5, Nb_2O_5, TiO_2)/B$, and then put into a graphite crucible with graphite lid. The whole assembly was placed into a vacuum furnace. Afterwards, the system was heated at a rate of 8 K/min to the desired temperatures and maintained for 1 h, followed by the furnace cooling to room temperature. The whole heating and cooling process was carried out in flowing Argon gas (99.99%, purity) with flowing rate of 200 sccm.

To analyze the formation mechanism of HMD-1 powders, the first-principles calculations based on density functional theory (DFT) were performed using Vienna Ab-Initio Simulation Package (VASP) [16,17]. The projected-augmented-waves with the Perdew-Burke-Ernzerhof exchange-correlation potentials were adopted [18,19]. It was already documented that the binary metal diborides, including $HfB_2$, $ZrB_2$, $TaB_2$, $NbB_2$ and $TiB_2$, possessed the similar crystal structures (hexagonal, space group of P6/mmm) [20,21]. Consequently, HMD-1 would maintain the same crystal structure after forming solid solutions with metal atoms randomly occupying 1a sites of equivalent point system in space group of P/6mmm. To build a chemical disordered multi-component metal diborides, the special quasi-random structure (SQS) approach was applied to build the supercell with 60 atoms by using the Alloy Theoretic Automated Toolkit code [22,23]. For the energy calculations on the SQS supercell, the plane-wave cutoff energy of 600 eV was used and the calculations were performed with a k-mesh grid of $2\pi \times 1/40$ Å$^{-1}$ for the supercell and $2\pi \times 1/60$ Å$^{-1}$ for the binary metal diborides with three-atom cell in VASP, respectively [24]. The electronic energy convergence criterion and the ionic force convergence criterion were $10^{-6}$ eV and 0.01 eV/Å, respectively. The equilibrium DFT energies and lattice parameters were then obtained at 0 K and 0 Pa.

The as-synthesized powders were analyzed by X-ray diffraction (XRD, X'pert PRO; PANalytical, Almelo, Netherlands), scanning electron microscopy (SEM, Supra-55; Zeiss, Oberkochen, Germany) equipped with energy dispersive spectroscopy (EDS) and transmission electron microscopy (TEM, Tecnai F30G2; FEI,

Eindhoven, Netherlands) equipped with EDS.

XRD characterization was first conducted to analyze the phase composition of the as-prepared powders at various temperatures. To observe the weak diffraction peaks more clearly, X-ray data was plotted on a logarithmic scale, as displayed in Fig. 1. When the synthesis temperature was 1773 K, it was found that the as-synthesized products were composed of two different solid-solution phases of metal diborides, namely, a dominant (Hf, Zr, Ta, Nb, Ti)$B_2$ phase and a minor (Hf, Zr)$B_2$ phase (JCPDS card no. 65-8680). With the increase of the synthesis temperature, these two solid-solution phases gradually merged and finally formed a single (Hf$_{0.2}$Zr$_{0.2}$Ta$_{0.2}$Nb$_{0.2}$Ti$_{0.2}$)$B_2$ phase with a hexagonal crystal structure of metal diborides without other phases at 1973 K. Therefore, the high-purity HMD-1 powders can be successfully synthesized by borothermal reduction method at 1973 K.

Fig. 2(a) presents SEM image of the as-synthesized HMD-1 powders. It can be clearly seen that the as-synthesized powders involve a large number of the sphere-like particles with the relatively uniform particle size of several hundred nanometers. Furthermore, the compositional uniformity of the as-synthesized powders (labeled by a dotted white square in Fig. 2(a)) is detected by EDS compositional maps at micrometer scale, as displayed in Fig. 2(b). The results show that the distribution of Hf, Zr, Ta, Nb and Ti elements is very homogeneous at micrometer scale without evident localization of any metal elements. This indicates that the as-synthesized HMD-1 powders possess the high compositional uniformity at micrometer scale.

Fig. 3(a) shows a typical bright-field TEM image of the as-synthesized HMD-1

powders. It can be clearly seen that the as-synthesized powders involve numerous individual nanocrystalline particles. Generally, TEM images can be used to measure the particle size. As a result, the particle sizes of the as-synthesized powders are measured for 50 individual particles. A Gaussian fitting to these data yields an average particle size of ~ 310 nm (Fig. 3(b)). Fig. 3(c) represents a representative selected area electron diffraction (SAED) pattern along zone axis $[12\bar{1}]$ of the as-synthesized powders. It obviously exhibits that the as-synthesized powders possess a single crystal hexagonal structure of metal diborides due to the well-arranged diffraction spots with the symmetry. High-resolution transmission electron microscopy (HRTEM) image (Fig. 3(d)) represents that the as-synthesized powders have a periodic lattice structure with a set of fringes with the d-space of 0.269 nm, corresponding to the {100} plane of $(Hf_{0.2}Zr_{0.2}Ta_{0.2}Nb_{0.2}Ti_{0.2})B_2$ phase, which is consistent with the calculated value (2.693 Å) from XRD pattern. In addition, an amorphous $B_2O_3$ layer of 2 ~ 3 nm can be observed on the surface of the as-synthesized powders (Fig. 3(d)), which can account for the presence of B element and a small amount of O element detected by EDS analyses (Table 1). In addition to B and O elements, five metal elements of Hf, Zr, Ta, Nb and Ti also can be detected by EDS, as displayed in Table 1. Obviously, their corresponding atomic percentages are the equal atomic quantity, which further confirm that the as-synthesized powders are composed of $(Hf_{0.2}Zr_{0.2}Ta_{0.2}Nb_{0.2}Ti_{0.2})B_2$ phase. Fig. 3(e) shows the scanning transmission electron microscopy (STEM) image and the EDS compositional maps of the as-synthesized powders at nanometer scale. Obviously, the distribution of Hf, Zr, Ta, Nb and Ti elements is highly uniform at nanoscale and no segregation or aggregation is found throughout the as-synthesized powders. In summary, combined to the results from XRD, SEM and TEM characterizations, the high-purity and superfine HMD-1 powders with high compositional uniformity from nanoscale to microscale have been successfully synthesized by borothermal reduction method at 1973 K.

To understand the formation mechanisms and chemical reactions associated to the synthesis of HMD-1 powders at 1973 K, the thermodynamics analysis of the possible chemical reactions is performed and the corresponding results are shown in Fig 4(a). In our case, the reaction precursors mainly consist of $HfO_2$, $ZrO_2$ $TiO_2$, $Nb_2O_5$, $Ta_2O_5$ and amorphous B powders. Accordingly, the possible reactions as well as the correlations between the standard Gibbs free energy ($\Delta G_{R,T}^{\theta}$) of these reactions and the temperature ($T$) can be described as follows:

$$HfO_2 + \frac{10}{3}B \rightarrow HfB_2 + \frac{2}{3}B_2O_3(l) \tag{1}$$

$$\Delta G_{R,T}^{\theta} = -78878 - T$$

$$ZrO_2 + \frac{10}{3}B \rightarrow ZrB_2 + \frac{2}{3}B_2O_3(l) \tag{2}$$

$$\Delta G_{R,T}^{\theta} = -77553 - 11T$$

$$TiO_2 + \frac{10}{3}B \rightarrow TiB_2 + \frac{2}{3}B_2O_3(l) \tag{3}$$

$$\Delta G_{R,T}^{\theta} = -223930 + 4T$$

$$\frac{1}{2}Nb_2O_5 + \frac{11}{3}B \rightarrow NbB_2 + \frac{5}{6}B_2O_3(l) \tag{4}$$

$$\Delta G_{R,T}^{\theta} = -405640 + 5T$$

$$\frac{1}{2}Ta_2O_5 + \frac{11}{3}B \rightarrow TaB_2 + \frac{5}{6}B_2O_3(l) \tag{5}$$

$$\Delta G_{R,T}^{\theta} = -240372 + 2T$$

$$HfB_2 + ZrB_2 + TaB_2 + NbB_2 + TiB_2 \rightarrow 5(Hf_{0.2}Zr_{0.2}Ta_{0.2}Nb_{0.2}Ti_{0.2})B_2 \quad (6)$$

$$\Delta G_{R,T}^{\theta} = 5\Delta G_{R,T}^{M}$$

$$ZrO_2 + HfO_2 + TiO_2 + \frac{1}{2}Ta_2O_5 + \frac{1}{2}Nb_2O_5 + \frac{52}{3}B \rightarrow$$

$$5(Hf_{0.2}Zr_{0.2}Ta_{0.2}Nb_{0.2}Ti_{0.2})B_2 + \frac{11}{3}B_2O_3 \quad (7)$$

$$\Delta G_{R,T}^{\theta} = -1026373 - T + 5\Delta G_{R,T}^{M}$$

where $\Delta G_{R,T}^{M}$ is the mixing Gibbs free energy of HMD-1, which can be calculated by the following equation:

$$\Delta G_{R,T}^{M} = \Delta H_{mix} - T\Delta S_{mix} \quad (8)$$

where $\Delta H_{mix}$ is the mixing enthalpy of HMD-1 and $\Delta S_{mix}$ is the mixing entropy of HMD-1. Owing to its insensitive to the temperature [25], the mixing enthalpy of HMD-1 can be estimated by the mixing enthalpy of HMD-1 at 0 K ($\Delta H_{mix}^{0K}$), which can be calculated by the following equation:

$$\Delta H_{mix}^{0K} = E_{HMD-1} - (E_{HfB_2} + E_{ZrB_2} + E_{TaB_2} + E_{NbB_2} + E_{TiB_2})/5 \quad (9)$$

where $E$ are the energies of the different systems after relaxation at 0 K and 0 Pa. The equilibrium lattice parameters of HMD-1 and the five metal diborides (HfB$_2$, ZrB$_2$, TaB$_2$, NbB$_2$ and TiB$_2$) at 0 K are first calculated by the first principles, as listed in Table 2. According to the calculated lattice parameters, the energies of HMD-1 and the five binary metal diborides at 0 K are then calculated by the first principles, as displayed in Table 2. Finally, using the energies from DFT calculations, the mixing enthalpy of HMD-1 is calculated to be about -0.232 kJ/mol by the Equation (9). In addition, the mixing entropy of HMD-1 can be expressed as [26]:

$$\Delta S_{mix} = -\frac{R}{3}\sum_{i=1}^{N} x_i \ln(x_i) \quad (10)$$

where $R$ is the ideal gas constant, $N$ is the element species in the metal sublattice of HMD-1, and $x_i$ is the molar fraction of the $i$th element in the metal sublattice of HMD-1. According to the Equation (10), the mixing entropy of HMD-1 is calculated to be about 0.536R. As a result, the standard Gibbs free energy of the reactions (6) and (7) can be mathematically expressed as:

$$\Delta G^{\theta}_{R,T} = -1160 - 23T \tag{11}$$

$$\Delta G^{\theta}_{R,T} = -1027533 - 24T \tag{12}$$

From Fig. 4(a), it's clear that the standard Gibbs free energies of all the reactions ((1) ~ (6)) are negative ($\Delta G^{\theta}_{R,T} < 0$), and thereby they can all proceed spontaneously. Nevertheless, it's worth noting that the reaction (7) has highest driving force among all those reactions as a result of its lowest standard Gibbs free energy. Among all those reactions, the reaction (6) is the one that has the lowest driving force due to its highest standard Gibbs free energy. XRD experimental results (Fig. 1) show that no any simple binary diboride products (e.g., $TiB_2$, $ZrB_2$) are detected in the as-synthesized powders at 1973 K except for HMD-1 product, suggesting only reaction (7) occurs to generate HMD-1 in the system while the other reactions ((1) ~ (6)) do not actually proceed. In addition, the synthesis experiment of HMD-1 powders is also conducted at 1973 K for 10 min and XRD result is shown in Fig. 4(b), from which no any simple binary diboride products (e.g., $TiB_2$, $ZrB_2$) are detected except for a (Hf, Zr, Ta, Nb, Ti)$B_2$ product and the unreacted metal oxides. This confirms that only reaction (7) occurs to generate HMD-1 in the system while the other reactions ((1) ~ (6)) do not actually proceed. In consequence, the high-purity HMD-1 powders can be successfully synthesized via a borothermal reduction method from thermodynamic aspect.

In summary, we had successfully synthesized the high-purity and superfine HMD-1 powders using a simple borothermal reduction method for the first time. The as-synthesized powders had an average particle size of ~ 310 nm and possessed a single-crystalline hexagonal structure at the same time. More importantly, they exhibited high compositional uniformity from nanoscale to microscale. In addition, on the basis of the first principles calculations, we demonstrated the formation mechanism of the as-synthesized HMD-1 powders from a thermodynamics aspect. This work will open up a new research for synthesizing high-entropy metal diboride powders.

**Acknowledgements**

We acknowledge financial support from the National Key Research and Development Program of China (No. 2017YFB0703200), Young Elite Scientists Sponsorship Program by CAST (No. 2017QNRC001), and National Natural Science Foundation of China (No. 51802100).

**Table 1.** The elemental atomic percentage of the as-synthesized HMD-1 powders by EDS analyses.

| Elements | Hf | Zr | Ta | Nb | Ti | B | O |
|---|---|---|---|---|---|---|---|
| At. % | 6.22 | 6.37 | 6.58 | 6.64 | 6.15 | 64.57 | 3.47 |

**Table 2.** Calculated equilibrium lattice parameters and DFT energies of the different systems at 0 K and 0 Pa.

| Systems | HMD-1 | $ZrB_2$ | $HfB_2$ | $TaB_2$ | $NbB_2$ | $TiB_2$ |
|---|---|---|---|---|---|---|
| Lattice parameters $a$ (Å) | 3.113 | 3.171 | 3.143 | 3.103 | 3.110 | 3.036 |
| Lattice parameters $c$ (Å) | 3.379 | 3.540 | 3.486 | 3.323 | 3.321 | 3.224 |
| Energies (eV/atom) | -8.589 | -8.326 | -8.801 | -9.053 | -8.609 | -8.144 |

**Figure captions**

**Fig. 1.** XRD patterns of the as-synthesized powders at different temperatures.

**Fig. 2.** SEM characterizations of the as-synthesized HMD-1 powders: (a) SEM image; (b) EDS elemental maps.

**Fig. 3.** TEM analysis of the as-synthesized HMD-1 powders: (a) TEM image; (b) histogram of the particle sizes with a Gaussian fitting to the data. The Gaussian peak is centered at 310 nm; (c) SAED pattern; (d) HRTEM image; (e) STEM image and the corresponding EDS compositional maps.

**Fig. 4.** Formation mechanism analysis of the as-synthesized HMD-1 powders: (a) Thermodynamics analysis of the possible chemical reactions among the precursors in the system; (b) XRD pattern of the as-synthesized products at 1973 K for 10 min.

**Figures**

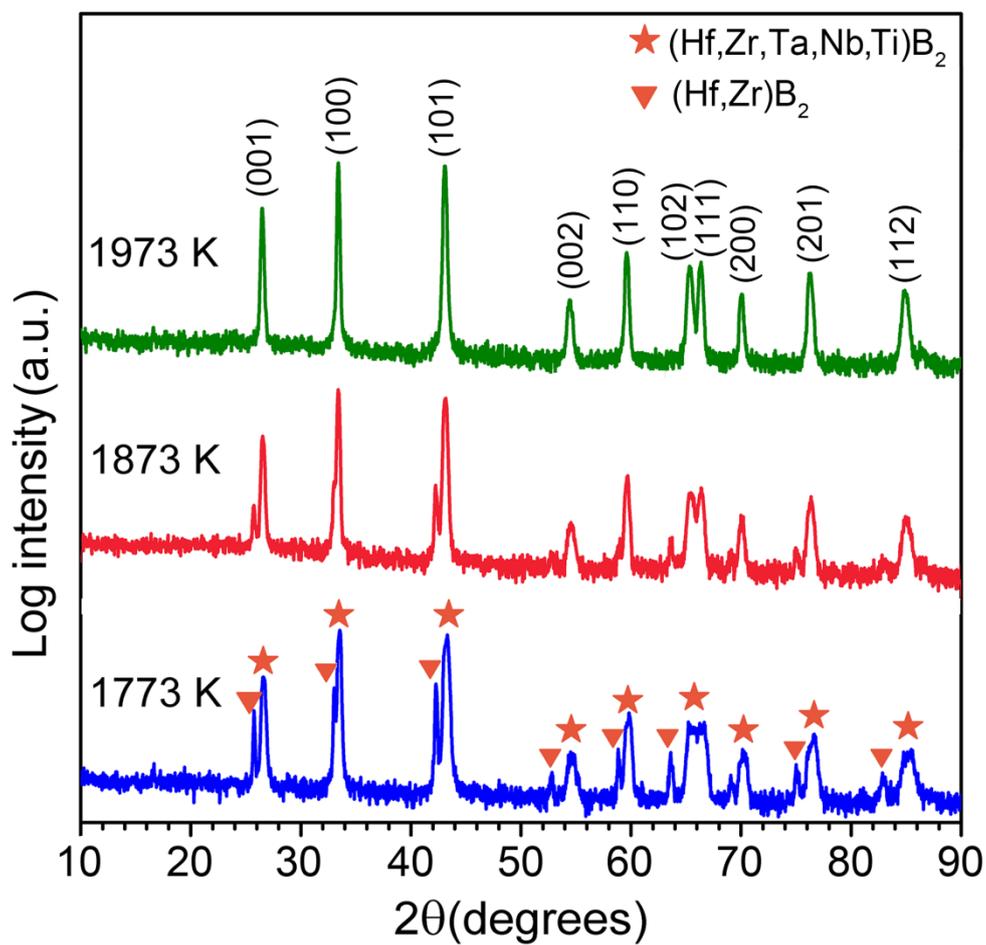

**Fig. 1.** XRD patterns of the as-synthesized powders at different temperatures.

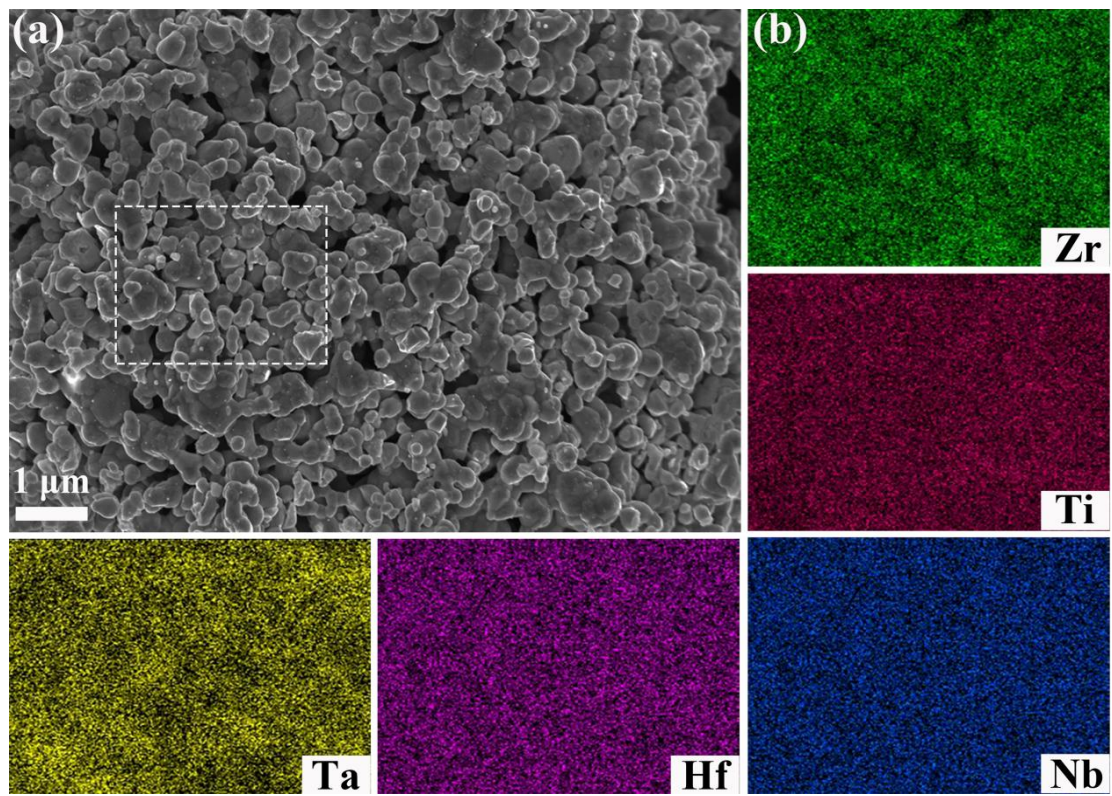

**Fig. 2.** SEM characterizations of the as-synthesized HMD-1 powders: (a) SEM image; (b) EDS elemental maps.

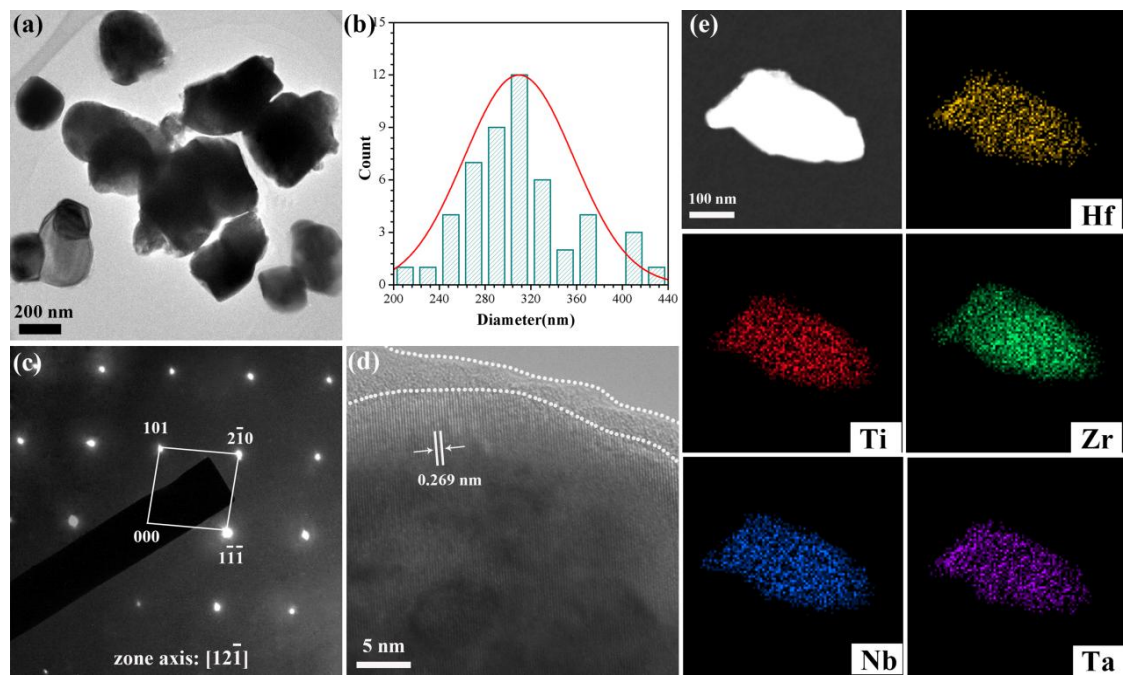

**Fig. 3.** TEM analysis of the as-synthesized HMD-1 powders: (a) TEM image; (b) histogram of the particle sizes with a Gaussian fitting to the data. The Gaussian peak is centered at 310 nm; (c) SAED pattern; (d) HRTEM image; (e) STEM image and the corresponding EDS compositional maps.

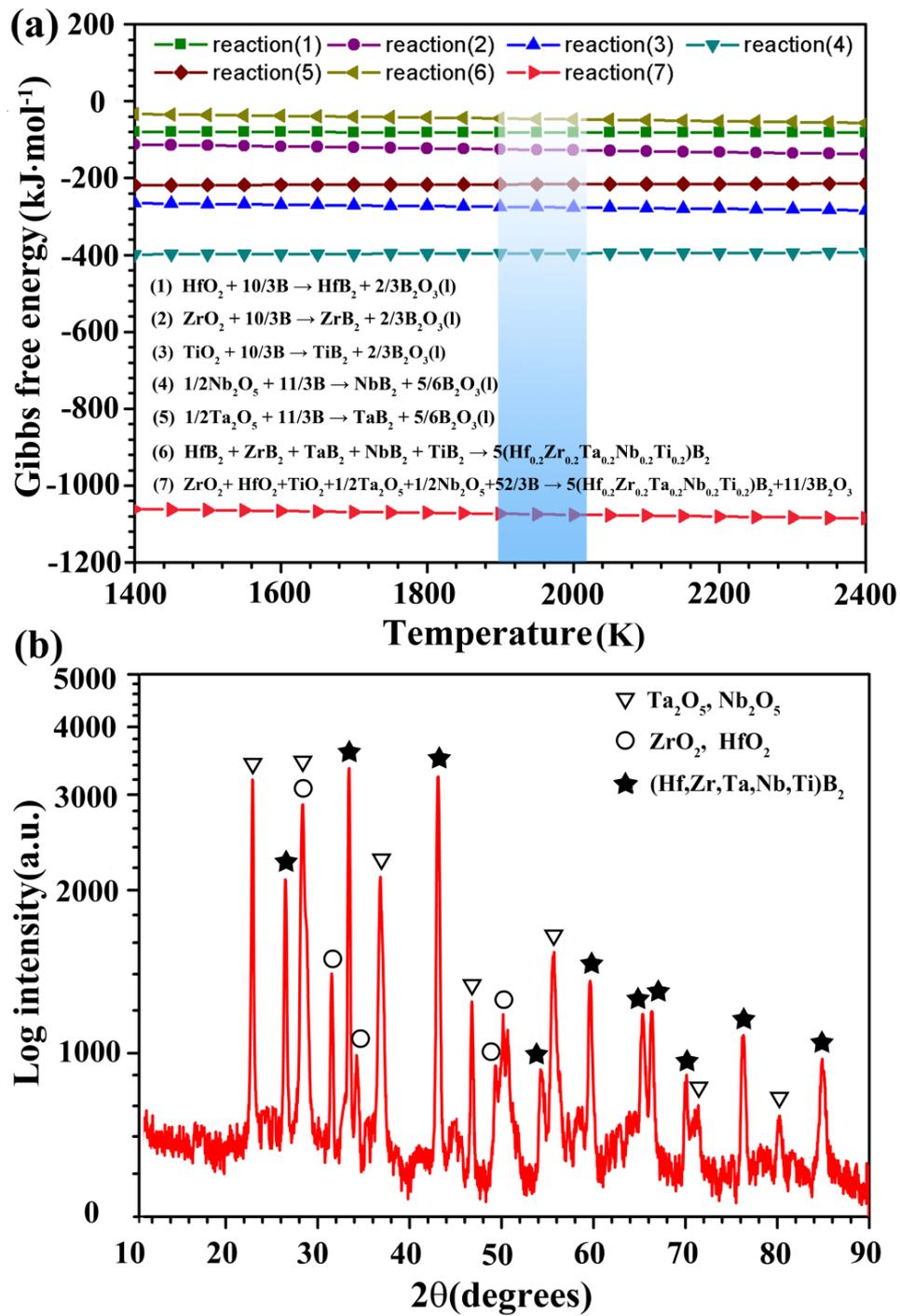

**Fig. 4.** Formation mechanism analysis of the as-synthesized HMD-1 powders: (a) Thermodynamics analysis of the possible chemical reactions among the precursors in the system; (b) XRD pattern of the as-synthesized products at 1973 K for 10 min.